\documentclass[prl,twocolumn,showpacs]{revtex4}
\usepackage{graphicx}
\usepackage{amsmath}
\newcommand{\n}{\nonumber}
\newcommand{\bn}{\begin{eqnarray}}
\newcommand{\en}{\end{eqnarray}}
\newcommand{\eml}{\end{multline}}
\newcommand{\bml}{\begin{multline}}

\begin{document}

\title {Mesoscopic Transport and Interferometry with Wavepackets of Ultracold atoms: Effects of
Quantum Coherence and Interactions}

\author{Kunal K. Das }
 \affiliation{Department of Physical Sciences, Kutztown University of Pennsylvania, Kutztown, Pennsylvania 19530, USA}

\date{\today }
\begin{abstract}
We propose a way to simulate mesoscopic transport processes with
counter-propagating wavepackets of ultracold atoms in quasi
one-dimensional (1D) waveguides, and show quantitative agreement
with analytical results. The method allows the study of a broad
range of transport processes at the level of \emph{individual
modes}, not possible in electronic systems. Typically suppressed
effects of quantum coherence become manifest, along with the effects
of tunable interactions, which can be used to develop a simpler type of sensitive atom
interferometer.

\end{abstract}
\pacs{05.60.Gg,73.23.-b,03.75.Dg,67.85.-d} \maketitle

\date{\today }
 \maketitle

Atomtronics, or electronics with ultracold atoms, is an emerging
field with broad potential. Exploratory papers on the subject have
focussed on atomic replicas of electronic components like
transistors
\cite{Zoller-transistor,Holland-PRL-atomtronics,Holland-PRA-atomtronics}.
The degenerate temperatures and the quantum nature of ultracold
atoms however, makes atomtronics akin to nanoscale mesocopic
processes \cite{Ferry-Goodnick}, rather than traditional
electronics. Therefore, besides component design, progress in
atomtronics calls for the study and simulation of the transport
mechanisms by which mesoscopic circuits operate.

Fermionic atoms in waveguides can mimic electrons in nanowires
\cite{Das-Aubin-PRL2009}. But it is the possibility of bosonic
carriers that makes atomtronics more than an imitation of mesoscopic
electronics. Carrier statistics can influence the characteristics of
atomtronics components \cite{Holland-PRA-atomtronics}, although not
their functionality. But in atomtronics implementations of
mesoscopic transport, bosons can offer significant advantages by
bypassing certain assumptions implicit in the solid state such as:
(i) multiple modes always present with fermionic carriers, that
suppress coherent correlations and (ii) fixed inter-particle
interactions, often ignored in models of ballistic transport in
nanowires \cite{Ferry-Goodnick}.

Transport experiments with degenerate bosons require a different approach, since there is no exclusion principle to guarantee non-vanishing momenta essential for fermionic transport processes. We therefore propose to simulate the basic Landauer \cite{Landauer,Ferry-Goodnick} paradigm of mesoscopic transport with wavepackets of cold bosons, which allows an  enhanced flexibility of operation that brings out features absent or suppressed in electronic transport.

\begin{figure}[t]
\includegraphics*[width=\columnwidth]{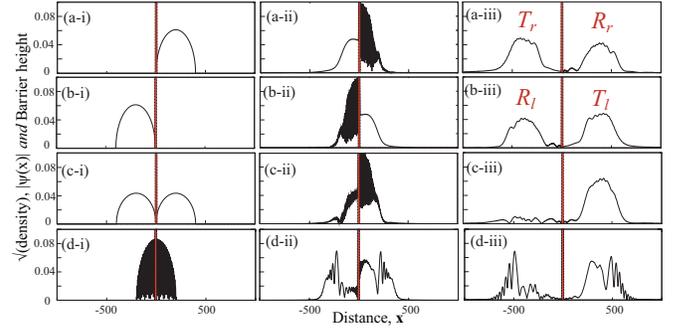}
\caption{Simulation of mesoscopic transport with wavepackets  of Thomas-Fermi profile $|\psi(x,0^-)|=\sqrt{b^2-(x-x_0)^2}, b=200$ and velocity $k=0.5$, for a static rectangular barrier slightly shifted from the origin: (a) Left incident packet, (b) Right incident packet, (c) Simultaneous left and right incident packets propagating coherently, and (d) Single wavepacket split into 50-50 superposition of $\pm\hbar k$ momentum states, similar to (c) but with additionally two outbound fractions.
}\label{Fig-1-snapshots}
\end{figure}

\emph{Mesoscopic Transport with wavepackets}: In mesoscopic solid state systems transport is described by quantum scattering \cite{Landauer,Ferry-Goodnick}.  Fermionic carriers move ballistically in quasi 1D nanoscale leads between macroscopic contacts that act as absorbing reservoirs for the carriers. Any device connected to the wire acts as a scatterer. In a single-channel circuit with two leads (left$\rightarrow l$, right $\rightarrow r$), the particle current for spin-polarized fermions (F),
\bn\label{fermion-current} J_F=\int_0^{\infty}\frac{dk}{2\pi}\left[v(k)f_l(k)T_l(k)-v(k)f_r(k)T_r(k)\right],\en
is determined by the Fermi distribution functions $f_{l(r)}$ of the
contacts, the transmission probabilities $T_{l(r)}$ for left and
right incident particles and their velocities $v(k)$. The underlying
picture is that of carriers injected at all available modes $k$ in
both leads and the net current is given by the weighted sum of the
current at \emph{each mode},
\bn\label{boson-current} J_B(k)=\frac{v}{2}[(f_l-f_r)+(f_lT_l+f_rR_r)-(f_lR_l+f_rT_r)],\en
termed bosonic (B) being single mode, in contrast with
multi-mode fermionic currents.
The first term is due to inbound particles, bias driven with no
scatterer. The last two terms are \emph{incoherent}
sums of the  reflected and transmitted fractions, outbound from the
scatter.

The single mode current is determined by the scattering
probabilities, therefore it can be directly simulated with ultracold
atoms: Start with a wavepacket $\psi(x,t=0^-)$ of ultracold atoms
(of axial extent $2b$)  in a quasi-1D harmonic trap
\cite{Das-Aubin-PRL2009}, first on the left and then on the right of
the ``device" (a scattering potential implemented with tightly
focussed lasers, blue or red detuned for barriers or wells). To
initiate the transport experiment, at $t=0$ the axial trap is turned
off and the atoms given an inward momentum $\pm \hbar k$ by using
Bragg beams \cite{Ketterle}, $\psi(x,0^+)=e^{\pm ikx}\psi(x,0^-)$.
The wavepacket is allowed to evolve for $t_f>2kb$ such that
the scattered wavepacket has little overlap with the device, and
then the spatial and momentum distribution imaged. The integrated
densities $n_\pm(t_f)=\int dk\theta(\pm k)|\psi(k,t_f)|^2$ of the
left(-) and right(+) moving scattered fractions directly give the
scattering probabilities; which, for broad packets,
match those of plane waves $e^{\pm ikx}$. Along with
$f_{l(r)}$ of the system, they
determine $J_B(k)$. Snapshots of the propagation of
the spatial wavefunction with a split-step operator method are shown
in Fig.~1(a,b).
\begin{figure}[t]
\includegraphics*[width=\columnwidth]{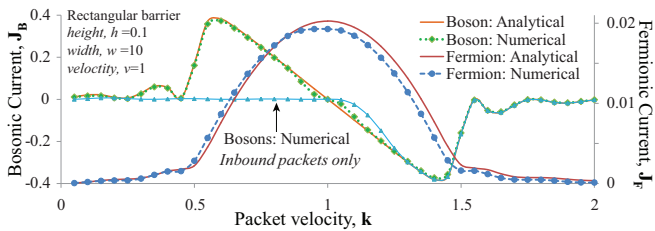}
\caption{Wavepacket simulation with a single split wavepacket as in Fig.~\ref{Fig-1-snapshots}(d), compared to exact analytical \cite{Das-Aubin-PRL2009} current profile of a `snowplow' quantum pump operating by a translating potential. Fermionic and Bosonic currents are plotted. Inbound packets alone [Fig.~\ref{Fig-1-snapshots}(c)] are insufficient for $k<v$. Numerical curves are interpolations of the marked computed values.
}\label{Fig-2-snowplow}
\end{figure}
Fermionic transport is simulated
by  replacing the integral  in Eq.~(\ref{fermion-current}) by a
Riemann sum sampled at discrete intervals of $\Delta k$:
\bn J_F=\int_{-\infty}^{\infty} \frac{dk}{2\pi}f(k)J_B(k)\simeq
\frac{\Delta k}{2\pi}\sum_{i}f(k_i)J_B(k_i).\en
Figure \ref{Fig-2-snowplow} shows the accuracy of this approach for
both single mode bosonic transport and integrated fermionic
transport for a  `snowplow' quantum pump \cite{Das-Aubin-PRL2009} of interest in mesoscopic
physics. The key point is that $J_B$ needs be sampled only at some
points in $k$-space to map out $J_F$.

The power of the method is in its simplicity and the variations it
allows for exploring transport features, many not possible in
mesocscopic systems: (i) Nonlinear transport, tunable by Feshbach
resonances \cite{Grimm-RMP-Feshbach} in packets of BEC in an optical
dipole trap \cite{Grimm-optical-dipole}; (ii) quantum to
semi-classical limit by narrowing packet widths; (iii) transport and
resonance transmission with different potentials by sculpting laser
profiles and bias fields; (iv) time-varying potentials; (v)
coherence effects by propagating left and right going packets
simultaneously [Fig.~\ref{Fig-1-snapshots}(c)]; (vi) adjustable
periodic potential with optical lattices.

\emph{Simulation and Physical Parameters}: To describe coherent superpositions and
velocity-changing time-varying scatterers, Eq.~(\ref{boson-current}) is
generalized to
\bn J[\psi(k,t)]=(\hbar/m) \langle
\psi(k,t)|k|\psi(k,t)\rangle/\langle \psi(k,t)|\psi(k,t)\rangle.\en
This, at $t=t_f$ after scattering, corresponds to the last two terms of Eq.~(\ref{boson-current}); the first term determined by the Fermi functions, vanishes for biasless
transport $f_l=f_r$.  Possible initial states in position space $\psi(x,0^-)$ are shown in Fig.~\ref{Fig-1-snapshots}. For incoherent sum of left  [Fig.~\ref{Fig-1-snapshots}(a)] and right going [Fig.~\ref{Fig-1-snapshots}(b)] packets, the currents are computed separately and averaged $J=\frac{1}{2}[J[\psi_l(k,t)]+J[\psi_r(k,t)]]$.  Packets are weighted by $f_{l(r)}$ for biased transport.

Assuming degenerate bosons, we use initial Thomas-Fermi profile
$|\psi(x,0^-)|^2=b^2-(x-x_0)^2$. The packet(s) are propagated with
the 1D Gross-Pitaevskii (GP) equation
$ [-\frac{\hbar^2}{2m}\partial_x^2+V(x)+g_{1D}|\psi|^2]\psi=-i\hbar\partial_t\psi\label{gross-pitaevskii},$
the nonlinearity \cite{das-crossover} measured by the effective 1D
interaction $g_{1D}=2aN$ ($a\rightarrow$ scattering length,
$N\rightarrow$ number of atoms). The transverse trap frequency
$\omega_r$ defines our units $l=\sqrt{\hbar/(m\omega_r)},
E_0=\hbar\omega_r$ and $\tau=\omega_r^{-1}$.  In the non-interacting
case $g_{1D}=0$, but the Thomas-Fermi profile is still used, as
results are insensitive to packet shape, if wide enough to
approximate plane waves [Fig.~\ref{Fig-6-turnstile}]. Also, a small
nonlinearity $g_{1D}\simeq 1$ can substantially broaden the initial
packet and still approximate linear behavior.

Any atom species with a BEC and a Feshbach resonance through zero
scattering length may be used. For example, with $^{39}K$ in a trap
of radial and axial frequencies of $\omega_r=2\pi\times 600$ Hz and
$\omega_a=2\pi\times0.6$ Hz, possible in current experiments
\cite{Inguscio-K39}, our units are $l=0.65 \mu$m, $E_0=0.029 \mu$K
and $\tau=0.26$ ms. With scattering length tuned to $0.05 a_B$, in
these units $g_{1D}=0.81$ and packet width $b=106$ for $N=10^5$
atoms \cite{das-anisotropic}, appropriate for testing linear
transport. A scattering length of $1.5 a_B$ gives $g_{1D}=24$ and $b=330$, sufficient to test nonlinear transport described here. Our simulations
use wider packets $b=600$ only for precise matching of plane wave results [Fig.~\ref{Fig-6-turnstile}]. Typical packet velocity $l/\tau=2.5 {\rm mm/s}$ defines
the time scale of experiments, $2kb\simeq 50$ ms.

\begin{figure}[t]
\includegraphics*[width=\columnwidth]{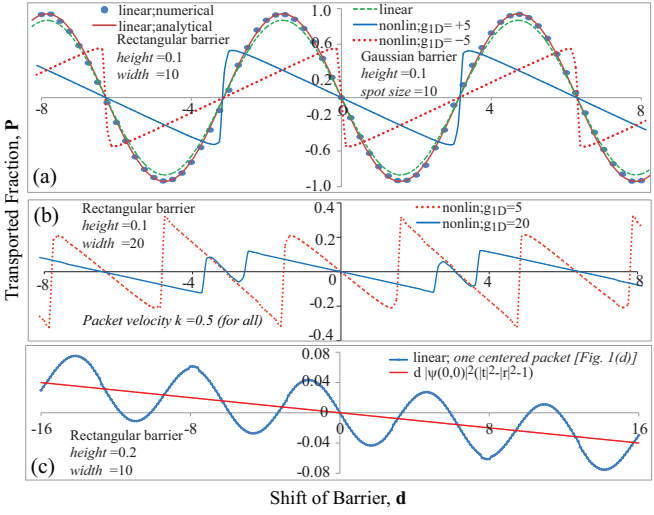}
\caption{
Coherent transport by a shifted static barrier: (a) Two-wavepacket simulation [Fig.~\ref{Fig-1-snapshots}(c)] for rectangular barrier (dots) exactly matches analytical curve. Simulation with Gaussian barrier (laser profile) is close.  Also shown are effects of (a) positive and negative nonlinearity $g_{1D}$ and (b) different magnitudes of positive nonlinearity. (c) For a centered packet [Fig.~\ref{Fig-1-snapshots}(c)] small initial imbalance, due to barrier shift, tilts the net transport. \emph{Numerical curves are interpolations of points at intervals marked by the `dots' in} (a).
}\label{Fig-3-1barrier-shift}\vspace{-5mm}
\end{figure}

\begin{figure}[b]
\includegraphics*[width=\columnwidth]{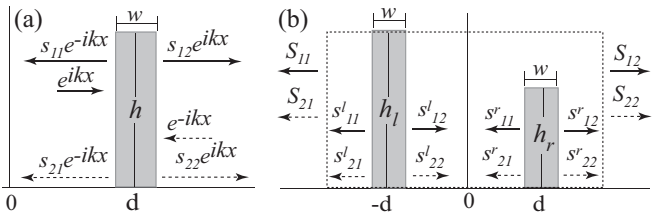}
\caption{Scattering matrix elements: (a) Single barrier shifted from
the origin,(b) Double-barrier potential of separation $2d$.
}\label{Fig-4-potentials}
\end{figure}

\emph{Transport Interferometry with Static Potentials}: Mesoscopic transport assumes lack of phase coherence among individual carriers, due to the randomization in the reservoirs \cite{Ferry-Goodnick}; hence the \emph{incoherent} sum of transmission probabilities in Eq.~(\ref{fermion-current}). With trapped atoms we can relax this condition with some interesting consequences.

Consider a symmetric \emph{static} scatterer in 1D with
no potential gradient [Fig.\ref{Fig-4-potentials} (a)]. Two identical wavepackets of momenta $\pm \hbar k$
simultaneously incident on opposite sides, should not give net current since scattering probabilities are
independent of the side of incidence. That is indeed so if
the scatterer is centered at the origin. But if it is shifted a distance $d$ from the origin, we
observe net flow, as shown in
Fig.~\ref{Fig-1-snapshots}(c). The transport fraction $P=n_+-n_-$
depends sinusoidally on the shift $d$, as shown in Fig.~\ref{Fig-3-1barrier-shift}(a).

\begin{figure}[t]
\includegraphics*[width=\columnwidth]{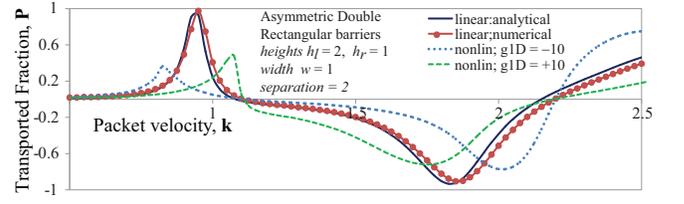}
\caption{Coherent transport by a static asymmetric double barrier
potential: Wavepacket simulation for linear and nonlinear
propagation; linear case matches analytical curve. \emph{Numerical
plots are interpolations at intervals marked on one}.
}\label{Fig-5-2barrier-shift}
\end{figure}
Classically impossible, such a current is a purely quantum effect
due to coherent superposition of the left and right going packets;
underscored by the fact that it is zero if the packets are incident
\emph{separately} as in Fig.~\ref{Fig-1-snapshots}(a) and (b), and then
the resulting currents are added.

This can be understood by considering the scattering matrix elements [Fig.~\ref{Fig-4-potentials}] of plane waves:
$ s_{12}=s_{21}=t; s_{11}=r^{2ikd};s_{22}=re^{-2ikd}$.
The transmission amplitudes are unaffected by the shift. But the
reflection amplitudes undergo phase shifts, so the coherent sums on
both sides, $\psi_\pm(x)=(t+re^{\mp2ikd})e^{\pm ikx}$ give:
\bn N={\textstyle \frac{1}{2}}[|\psi_+(x)|^2+|\psi_-(x)|^2]
&=&1+2\cos(2kd){\rm Re}\{t^*r\}\n\\P={\textstyle \frac{1}{2}}[|\psi_+(x)|^2-|\psi_-(x)|^2]&=&2\sin(2kd){\rm Im}\{t^*r\}\label{shift-equation}\en
Number conservation requires ${\rm Re}\{t^*r\}=0$, but generally
${\rm Im}\{t^*r\}\neq 0$ yielding non-vanishing P. Even with no
shift, asymmetric scatterers can generate differential
reflection phases, leading to net flow. This is demonstrated in Fig.~\ref{Fig-5-2barrier-shift}, for the asymmetric double-barrier configuration of Fig.~\ref{Fig-4-potentials}(b). While the transmission amplitudes are side-symmetric, the reflection
amplitudes
$ S_{11(22)}= r_{l(r)}e^{-2ikd}+t_{l(r)}^2r_{r(l)}e^{+2ikd}/(1-r_rr_le^{+4ikd})$
differ in phase unless the two barriers are identical.

This effect is due to coherent superposition of the
two waves $\pm k$ introducing a spatial periodicity that breaks translational invariance. Effect of bias is different, as can be seen with a packet initially centered at $x=0$ [Fig~\ref{Fig-1-snapshots}(d)]. A barrier shift means more of the packet on one side. For $d\ll b$ , this adds a term in Eq.~(\ref{shift-equation}): $P\simeq d|\psi(0,0)|^2(|t|^2-|r|^2-1)+\sin(2kd){\rm Im}\{t^*r\}$ causing a linear tilt in P [Fig~\ref{Fig-3-1barrier-shift}(c)]. Here, P is reduced by $1/2$ as half of a centered packet is outbound.

Although non-classical, this is consistent with thermodynamics as
there is no net current in a thermal mixture. Even for a 50-50
mixture of orthonormal states $\cos(kx)$ and $\sin(kx)$ the net
current vanishes, as seen for the representative case of a shifted
$\delta$-potential $U\delta(x-d)$:
$J[\cos(k)]=-J[\sin(k)]=-\sin(2kd)\hbar k^2U/[2m(k^2+U^2)]$.

\emph{Coherence in time-dependent phenomena}: By running left and
right going packets separately or simultaneously the
role of quantum coherence in transport phenomena can be evaluated. We illustrate
with two different time varying potentials associated with the
mesoscopic process called quantum pumps \cite{Das-Aubin-PRL2009}:
(i) \emph{snowplow}, where a single potential barrier
[Fig.~\ref{Fig-4-potentials}(a)] translates at uniform velocity
$d=vt$ and (ii) \emph{turnstile}, where heights of two barriers
[Fig.~\ref{Fig-4-potentials}(b)] vary out of phase,
$h_l=h(1+\sin(\omega t))$ and $h_r=h(1+\cos(\omega t))$.  For the
snowplow, there is absolutely no difference in the current profile
[Fig.~\ref{Fig-2-snowplow}] whether the two packets are run
simultaneously or separately. This is because the shift $d$ is now a
function of time, so the sinusoidal dependence on the shift averages
out.  But, the turnstile pump shows a significant difference whether
the left and right going packets interfere coherently or not, as
seen in Fig.~\ref{Fig-6-turnstile}(a). This supports our earlier
conclusions \cite{Das-Aubin-PRL2009} that the snowplow pump can be
simulated classically, but turnstile pumps involve quantum
interference.

\begin{figure}[t]
\includegraphics*[width=\columnwidth]{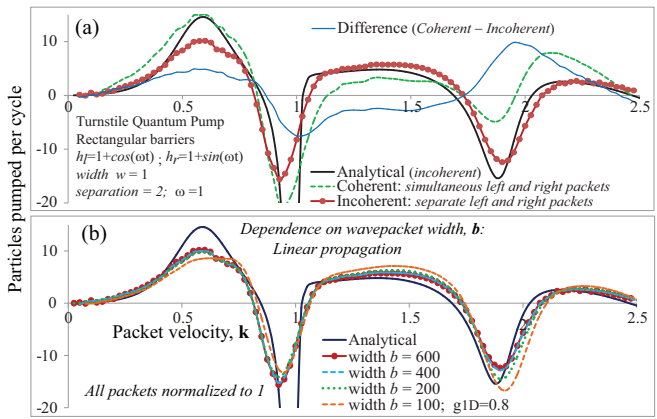}
\caption{Wavepacket simulation of a turnstile pump: (a) Coherent and incoherent transport compared to analytical results [3]. (b) Convergence with packet width ($b=400$ and $600$ indistinguishable); our estimates for $^{39}K$ correspond to $b=100$, the small nonlinearity has no perceptible effect.  \emph{Plots are interpolations of points at intervals marked on one.}
}\label{Fig-6-turnstile}
\end{figure}

\emph{Effects of Interactions}: Even a small interaction-induced
nonlinearity can lead to a dramatic change of the transport
features.  As shown in Fig.~\ref{Fig-3-1barrier-shift}  the
dependence of the transport fraction $P$ on the shift $d$ changes
from a sinusoidal to a triangular pattern, with sharp changes at
specific values of the shift. Plots for small \emph{positive} and
\emph{negative} nonlinearities are \emph{mirror images} of each
other across the anti-nodal planes of the  sinusoidal curves for the
corresponding linear case [Fig.~\ref{Fig-3-1barrier-shift}(a)]. Even
for the asymmetric barriers [Fig.~\ref{Fig-5-2barrier-shift}],
nonlinearity sharpens the drop-off of the first peak. Since the
superposition principle does not apply for nonlinear equations,
general analytical solutions may not be possible, even with
stationary solutions on hand \cite{Schlagheck-2007}. Therefore, our
method can be a valuable tool for probing nonlinear scattering and
transport, by enabling direct comparison of numerical simulations
and experiments in the same framework.

\begin{figure}[b]
\includegraphics*[width=\columnwidth]{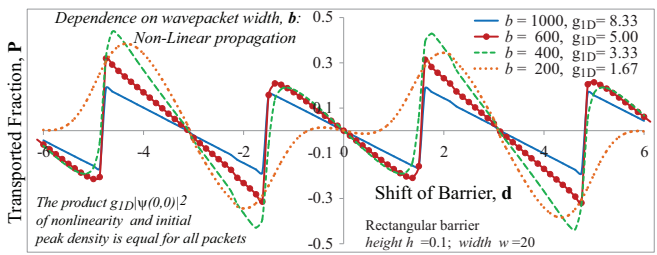}
\caption{Convergence with packet width for wavepacket simulation of
nonlinear transport, keeping $g_{1D}|\psi(0,0)|^2$ fixed; shown for
coherent transport due to a shifted barrier. \emph{Plots are
interpolations of points at intervals marked on one.}
}\label{Fig-7-nonlinear}
\end{figure}

Nonlinear propagation is sensitive to the packet shape since the nonlinear term in
the GP equation depends on $|\psi|^2$ which is
non-uniform for a wavepacket, unlike for a plane wave. Therefore,
rather than fix the normalization $\int dx |\psi(x,t)|^2$ as we vary the packet
widths (as done in the linear case), we fix the product $g_{1D}|\psi(0,0)|^2$, where $|\psi(0,0)|^2$ is the initial peak density. This leads to consistent convergence with the nodes and the turning
points occuring at convergent values, shown in Fig.~\ref{Fig-7-nonlinear}. This does not happen if $\int dx |\psi(x,t)|^2$ is fixed instead. Note our nonlinearity $g_{1D}|\psi(0,0)|^2\sim 0.02$ is small enough to be in a regime where GP approximation is valid \cite{Schlagheck-2010}.

\emph{Conclusions and Outlook}: We have presented an experimentally
feasible and theoretically accurate way to conduct mesoscopic
transport experiments relevant in solid state nanocircuits, with
ultracold atoms. The tunability of parameters and absence of Coulomb
interaction will allow study of transport phenomena with much
broader possibilities. Coherence effects inherently suppressed in
solid state systems can be made manifest with single mode transport
studies possible with atoms. The predictable and sensitive coherent
transfer due to small barrier shifts and asymmetries suggests applications for
sensitive atom interferometers \cite{Berman,Pritchard-RMP} where the
device laser is connected to a sensor; this can be quite robust
since it measures scattered densities well separated in position and
momentum space, without multi-step splitting-recombination of wavefronts,
typical of interference effects as in Mach-Zehnder
interferometers \cite{Pritchard-RMP}.

Nonlinearity in quantum coherent transport and scattering has very
rich behavior as our simulations indicate. Our approach provides a
simple numerical way, and a viable experimental method for probing
such effects,  still largely unexplored. Sharper variations with
nonlinearity, as in Fig.~\ref{Fig-3-1barrier-shift}, suggest that
small interactions could actually be used to enhance
interferometeric sensitivity.

This work was funded by an NSF grant PHY-0970012. We thank S. Aubin and T. Opatrn\'{y} for
useful discussions.

\vfill

\end{document}